\documentclass[aps,preprint]{revtex4}%
\usepackage[latin9]{inputenc}
\usepackage{mathtools}
\usepackage{amsmath}
\usepackage{amssymb}
\usepackage{graphicx}
\usepackage{xcolor}
\usepackage{xspace}
\usepackage{colortbl}
\usepackage{ragged2e}
\usepackage{rotating}
\usepackage{amsfonts}
\usepackage{graphics}%
\providecommand{\U}[1]{\protect\rule{.1in}{.1in}}
\makeatletter
\@ifundefined{textcolor}{}
{
\definecolor{BLACK}{gray}{0}
 \definecolor{WHITE}{gray}{1}
 \definecolor{RED}{rgb}{1,0,0}
 \definecolor{GREEN}{rgb}{0,1,0}
 \definecolor{BLUE}{rgb}{0,0,1}
 \definecolor{CYAN}{cmyk}{1,0,0,0}
 \definecolor{MAGENTA}{cmyk}{0,1,0,0}
 \definecolor{YELLOW}{cmyk}{0,0,1,0}
}
\graphicspath{{New_Constraints_muonic_helium_sw60_v3_graphics/}{New_Constraints_muonic_helium_sw60_v3_tcache/}{New_Constraints_muonic_helium_sw60_v3_gcache/}}
\DeclareGraphicsExtensions{.pdf,.eps,.ps,.png,.jpg,.jpeg}

\makeatother
\begin{document}
\title{The alpha particle charge radius, the radion and the proton radius puzzle}
\author{F. Dahia}
\email{fdahia@fisica.ufpb.br}
\affiliation{Departament of Physics, Universidade Federal da Paraíba Grande, João Pessoa -
PB, Brazil}
\author{A. S. Lemos}
\email{adiellemos@gmail.com}
\affiliation{Departamento de Física, Universidade Federal de Campina Grande, Caixa Postal
10071, 58429-900 Campina Grande, Paraíba, Brazil}

\begin{abstract}
Recent measurements of the Lamb shift of muonic helium-4 ions were used to
infer the alpha particle charge radius. The value found is compatible with the
radius extracted from the analysis of the electron-helium scattering. Thus,
the new spectroscopic data put additional empiric bounds on some free
parameters of certain physics theories beyond the Standard Model. In this
paper, we analyze the new data in the context of large extra-dimensional
theories. Specifically, we calculate the influence of the radion, the scalar
degree of freedom of the higher-dimensional gravity, on the energy difference
between the 2S and 2P levels of this exotic atom. The radion field is related
to fluctuations of the volume of the supplementary space, and, in comparison
with the tensorial degrees of freedom, it couples to matter in a different
way. Moreover, as some stabilization mechanism acts exclusively on the scalar
degree of freedom, the tensor and scalar fields should be treated as
phenomenologically independent quantities. Based on the spectroscopic data of
muonic helium, we find constraints for the effective energy scale of the
radion as a function of the alpha particle radius. Then, we discuss the
implications of these new constraints on the proton radius puzzle.

\end{abstract}
\maketitle

\section{Introduction}

At the end of the last century, interest in extra-dimensional theories was
renewed by braneworld theories. Their original formulation was proposed to
explain the discrepancy between the scales of the electroweak and
gravitational interactions \cite{add1,add2}. According to these models, our
ordinary universe is a $4$-dimensional hypersurface embedded in a
higher-dimensional space \cite{add1,add2,rs1,rs2}. This geometric
interpretation follows from the assumption that particles and fields of the
Standard Model are trapped to the brane and unable to escape to extra
dimensions, unless they were subjected to processes involving energy scales
far beyond TeV scale. Thus, the apparent four-dimensionality of spacetime
would be a consequence of the existence of confinement mechanisms that keeps
the particles and fields stuck in the brane \cite{add1,rubakov}.

Gravity, in contrast, has access to extra dimensions at energies even below
the weak scale. The spreading of the gravitational field in the additional
directions would be the reason why gravity appears to be much weaker than the
other interactions at large distances. In this way, the braneworld scenario
would provide a simple and alternative explanation for the hierarchy problem.

At the same time, these models also predict that the strength of the
gravitational interaction is greatly magnified at small length scales. This is
a very interesting feature because, due to this modification, the high
dimensionality of spacetime could, in principle, have measurable effects on
many phenomena that take place in the brane. This expectation has motivated
numerous researches in several areas of physics (such as high-energy particles
\cite{lhc}) aimed at probing empirical signals of extra dimensions, by
investigating the behavior of the gravitational field at short distances.

In a submillimeter scale, for instance, the inverse-square law of gravity has
been tested in laboratories using torsion balances. Usually, in these
laboratories experiments, the higher-dimensional gravitational potential of a
pointlike mass is parameterized by a power-law-like or Yukawa-like potentials
\cite{murata} depending on the probed distance $r$ from the source in
comparison to the compactification radius $R$. According to torsion balance
experiments, the extra dimension radius should satisfy the constraint
$R\leq44$ $\mu m$ \cite{hoyle01,hoyle04,hoyle07}, when the theoretical
deviation is expressed in the Yukawa parametrization.

More recently, relying on the high precision now achieved in atomic transition
measurements, hydrogen-like atoms have also been considered in the search for
deviations of Newton's law of gravitation at the Angstrom scale
\cite{atomicspec1,atomicspec2,atomicspec3,molecule,safranova,h,lemos3,lemos4,ryd,lemos5,pHebound,pHe}%
. In this regard and for the purpose of our later discussion, it is
interesting to mention that transitions involving $S$-level are not computable
in the thin brane model with two or more additional dimensions. This problem
can be avoided in thick brane scenarios, where constraints for corrections of
the inverse square law due to extra dimensions were obtained from the analysis
of the hydrogen atomic energy spectrum \cite{lemos1}. In the thick brane
framework, Standard Model particles are confined to the 3-brane, but their
wave-functions extend somewhat in the transverse direction over a range of
order of the brane thickness. In the thin brane models, the confinement is of
delta type \cite{add1,add2,rs1,rs2}.

The thick brane model has also been applied to study muonic hydrogen
spectroscopy in order to investigate the proton radius puzzle in the
extra-dimensional scenario. The proton radius puzzle is the incompatibility in
the measurement of the proton charge radius obtained from experiments
involving electron-proton interaction and muonic atom spectroscopy
\cite{proton}. This conundrum arose from the measurement of the proton radius
extracted through the $2S-2P$ Lamb shift of muonic atoms \cite{nature,science}%
. Many proposed theoretical models have attempted to explain the discrepancy
between the results, considering that this would be a possible indication of
an additional force beyond the Standard Model interactions
\cite{proton,new9,new10,li,wang,new11,new12}. In the extra-dimensional thick
brane scenario, according to Ref. \cite{lemos2}, the energy excess found in
the measurement of the $2S-2P$ transition using muonic hydrogen spectroscopy
could be accounted for by the modification of gravitational interaction.

The muonic atom is obtained in laboratory by replacing the electron with a
muon. As the muon is about 207 times heavier than the electron, the
gravitational interaction has a much greater effect on the energy levels of
this atom than on those of conventional hydrogen. In the thick brane scenario,
this magnification can be very impressive. For example, the amplification
factor can reach a figure of two billion in transitions involving the S-level
\cite{lemos2}, since, in this scenario, this factor is proportional to the
muon's gravitational mass multiplied by the atomic reduced mass raised to the
3rd power. Therefore, the spectroscopy of muonic atoms could be very useful in
providing important constraints for modifications of gravity in the atomic domain.

Recently, new data from muonic hydrogen-like atomic transitions have been
obtained. Trying to shed light on the proton radius puzzle, precise
measurements of the $2S-2P$ transitions in muonic helium-$4$ have been used to
determine the ion charge radius of the $\alpha$ particle \cite{nature2}. The
results obtained are compatible with the value extracted from e-He scattering,
and, thereby, the puzzle is not present here. In this work, we intend to
discuss these new data in the context of the thick braneworld model.

In general, the phenomenological viability of braneworld, such as the ADD
model, also depends on the stabilization of the volume of the supplementary
space \cite{stabilization,antoniadis,goldberg,chacko}. To ensure this, an
additional mechanism should act on this degree of freedom. From the brane
perspective, the volume of the extra space can be viewed as a scalar degree of
freedom of the higher-dimensional gravity. Its behavior is described by a
scalar field known as the radion. Since the stabilization mechanism does not
operate on the tensorial degrees of freedom of higher-dimensional gravity, it
is recommendable to examine the effects of the radion and gravitons separately
and to try to establish experimental bounds on each one independently whenever possible.

Another important aspect to consider is that radions and gravitons couple to
matter differently. In a pure ultra-relativistic regime, Standard Model fields
on-shell would not produce radions. As a result, tests of extra-dimensional
theories in high-energy colliders cannot probe the tensorial and scalar modes
with the same accuracy in tree-level processes \cite{colliders}.

In this article, we examine the effects of the radion on the energy level of
hydrogen-like atoms in the thick brane scenario. More specifically, in section
II, the set up of the thick brane is described and the Hamiltonian of the
nucleus-lepton interaction trough the radion exchange is explicitly
determined. Using this Hamiltonian, we estimate, in section III, the
corrections of the energy levels of the muonic helium due to the gravitational
interaction between the muon and alpha nucleus mediated by the radion. The
correction term depends on a free parameter of the model corresponding to an
effective energy scale of the radion. By using recent data of the transition
$2S-2P$ in the ionic muonic helium, we obtain experimental bounds for the
effective energy scale of the radion as a function of the radius of the alpha
particle. In the section IV, we reconsider the issue of the proton radius
puzzle taking into account this new constraints and discuss whether these new
bounds could exclude higher-dimensional gravity as a possible explanation for
the puzzle. Finally, in Sect. V, we present our last remarks.

\section{The radion and the atomic level in thick brane scenarios}

Alternative theories of gravity have been proposed in several different
contexts \cite{murata}.\ In higher-dimensional scenarios, a straightforward
modification of gravity is obtained by simply extending the Einstein-Hilbert
action to the whole space-time, including the additional directions. In the
ADD model \cite{add1}, for instance, the supplementary space has a topology of
a torus $T^{\delta}$ with $\delta$ spacelike dimensions, and the action is
given by:
\begin{equation}
S_{G}=\frac{c^{3}}{16\pi G_{D}}\int d^{4}xd^{\delta}z\sqrt{-\hat{g}}%
\hat{\mathcal{R}}\text{,}%
\end{equation}
where $x$ denotes intrinsic coordinates of the brane and $z$ represents
coordinates of the extra space. The Lagrangian depends on the scalar curvature
$\hat{\mathcal{R}}$ of the ambient space and $\hat{g}$ is the determinant of
the spacetime metric (here, we are adopting the signature $\left(
-,+,\ldots,+\right)  $). The gravitational constant of the higher-dimensional
space is $G_{D}$ and $c$ denotes the velocity of light in vacuum.

According to the ADD model \cite{add1}, the background state is characterized
by a flat spacetime that contains an extra space whose volume is given by
$\left(  2\pi R\right)  ^{\delta}$, where $R$ denotes the compactification radius.

In the presence of confined matter in the brane, the metric of the ambient
spacetime will be determined by equations with the same form of the Einstein's
equations. In the weak field limit, the higher-dimensional version of the
linearized Einstein's equations can be written as
\begin{equation}
\square\hat{h}_{AB}=-\frac{16\pi G_{D}}{c^{4}}\bar{T}_{AB}\text{,}
\label{eq_for_h}%
\end{equation}
for the tensor $\hat{h}_{AB}=\hat{g}_{AB}-\eta_{AB}$, which describes the
perturbations of the geometry with respect to the Minkowski metric $\eta_{AB}$
in the first order of $G_{D}$ (the capital Latin indices run from $0$ to
$3+\delta$). The above equation is valid in coordinate systems where the
condition $\partial_{A}\left(  h^{AB}-\frac{1}{2}\eta^{AB}h_{C}^{C}\right)
=0$ is satisfied. The operator $\square$ corresponds to the D'Alembertian
associated to the Minkowski metric and the source term is $\bar{T}%
_{AB}=\left[  T_{AB}-(\delta+2)^{-1}\eta_{AB}T_{C}^{C}\right]  $, defined from
$T_{AB}$, which is the energy-momentum tensor of the fields stuck in the
brane. Due to the confinement, it is assumed that $T_{AB}$ can be written as
\cite{colliders}:
\begin{equation}
T_{AB}\left(  x,z\right)  =\eta_{A}^{\mu}\eta_{B}^{\nu}T_{\mu\nu}\left(
x\right)  f\left(  z\right)  \text{,} \label{T}%
\end{equation}
in a length scale greater than the thickness of the brane. Here, the Greek
indices go from 0 to $3$. In the above expression, $T_{\mu\nu}\left(
x\right)  $ describes the effective distribution of energy and momentum of the
confined fields along the brane, while the function $f(z)$ is related to the
profile of the fields in the transversal directions. In zero-width brane,
$f(z)$ would be a delta-like distribution. However, in a thick-brane model,
$f\left(  z\right)  $ would be some normalized distribution very concentrated
around the brane.

Now let us consider the solutions of equation (\ref{eq_for_h}) in the ambient
space whose topology is $\mathbb{R}^{3}\times T^{\delta}$. In this context, to
take into account the compact topology of the extra dimensions, it is useful
to describe the $\delta$-torus as a quotient space of the Cartesian space
$\mathbb{R}^{\delta}$. Thus, the effect of the compact dimensions of the torus
$T^{\delta}$ on solutions of (\ref{eq_for_h}) can be simulated by means of
mirror images of the source. The localization of these images in
$\mathbb{R}^{\delta}$ is determined by the equivalence relation that defines
the quotient space. So, it follows that the solution of the equation
(\ref{eq_for_h}) in the given topology for static sources can be written as:
\begin{equation}
\hat{h}_{AB}\left(  X\right)  =\frac{\hat{G}_{D}}{c^{4}}\sum_{i}\left(
\int\frac{\bar{T}_{AB}\left(  X_{i}^{\prime}\right)  }{\left\vert
X-X_{i}^{\prime}\right\vert ^{1+\delta}}d^{3+\delta}X_{i}^{\prime}\right)
\text{,} \label{h}%
\end{equation}
where, for the sake of simplicity, we write $\hat{G}_{D}=[16\pi\Gamma
(\frac{\delta+3}{2})/(\delta+1)2\pi^{\left(  \delta+3\right)  /2}]G_{D}$, with
$\Gamma$ as the gamma function. The variable $X=$ $\left(  x,z\right)  $
represents the coordinates of points in the ambient space. For $i=0$, the
coordinate $X_{i=0}^{\prime}$ is the position vector of the real source inside
the thick brane, and each $X_{i}^{\prime}$ can be interpreted as the position
vector of the source's mirror image $i$ in the space $\mathbb{R}^{3+\delta}$.

At large distances from the source, the three-dimensional behavior of the
gravitational field is recovered in the brane, i.e., the components $\hat
{h}_{AB}$ behave as $\left\vert x\right\vert ^{-1}$, for $\left\vert
x\right\vert \gg R$ \cite{kehagias}. But to reproduce the predictions of
General Relativity, two additional conditions should be satisfied: the
higher-dimensional gravitational constant should be related to the Newtonian
constant $G$ according to the formula $G_{D}=\left(  2\pi R\right)  ^{\delta
}G$ \cite{add1,colliders}; and the volume of the supplementary space should be
stabilized at long distance.

This question is relevant here because fluctuations of the extra-space volume
have influence on the asymptotic behavior of gravitational potential. In fact,
particles confined to the brane couple to $\hat{h}_{\mu\nu}$, the induced
metric in that hypersurface. But, in order to reproduce the General
Relativity's predictions at large distances, particles should be effectively
coupled to another tensor, let us say $h_{\mu\nu}$ (without a hat), whose
source term is not $\bar{T}_{\mu\nu}$ that appears in equation (\ref{h}), but,
instead, is the reduced energy-momentum given by $\bar{T}_{\mu\nu}^{\left(
GR\right)  }\equiv\left(  T_{\mu\nu}-1/2\eta_{\mu\nu}T_{\gamma}^{\gamma
}\right)  $. The basic distinction between these two tensors is the
coefficient multiplying the trace $T_{\gamma}^{\gamma}$.

By comparing their respective sources term ($\bar{T}_{\mu\nu}$ and $\bar
{T}_{\mu\nu}^{\left(  GR\right)  }$), we can see that the difference between
$\hat{h}_{\mu\nu}$ and $h_{\mu\nu}$ is proportional to $\eta_{\mu\nu}$. Based
on these considerations, it is convenient to decompose the metric perturbation
tensor as: $\hat{h}_{\mu\nu}=h_{\mu\nu}+\phi\eta_{\mu\nu}$, where $h_{\mu\nu}$
is sourced by $\bar{T}_{\mu\nu}^{\left(  GR\right)  }$, while the field $\phi$
is sourced by the residual tensor $\bar{T}_{\mu\nu}-\bar{T}_{\mu\nu}^{\left(
GR\right)  }$, which is proportional to $\eta_{\mu\nu}$. Taking into account
the form of the energy-momentum tensor of confined fields (\ref{T}), we find
that:
\begin{equation}
\phi=\frac{\hat{G}_{D}}{c^{4}}\frac{\delta}{2\left(  \delta+2\right)  }%
\sum_{i}\int\frac{T_{\gamma}^{\gamma}\left(  x^{\prime}\right)  f\left(
z_{i}^{\prime}\right)  }{\left\vert \vec{X}-\vec{X}_{i}^{\prime}\right\vert
^{1+\delta}}d^{3+\delta}X_{i}^{\prime}\text{.} \label{radion}%
\end{equation}

This field, which is a scalar quantity under brane coordinate transformations,
is related to the trace in transversal directions of the metric perturbation.
Indeed, from equations (\ref{T}) and (\ref{h}), we can see that $\phi=\hat
{h}_{\hat{a}}^{\hat{a}}/2$ (where the index $\hat{a}$ refers to the extra
directions). Therefore, it describes fluctuations of the extra-space's volume
around its background value $\left(  2\pi R\right)  ^{\delta}$. As the volume
can be expressed in terms of the compactification radius, $\phi$ is called the
radion field \cite{stabilization}. At this point it may be useful to mention
that in some references only the zero-mode oscillation of the scalar field is
called radion. But here we are calling radion the field given by the equation
(\ref{radion}) which also depends on the extra-dimensional coordinates.

As we have already mentioned, to recover the known behavior of gravity at
large length scale, it is necessary that $\hat{h}_{\mu\nu}$ tends to
$h_{\mu\nu}$ in the limit $\left\vert x\right\vert \gg R$ . This condition
demands that $\phi$ should be suppressed asymptotically. Usually, this is
achieved by some mechanism that adds mass to the radion field
\cite{stabilization,antoniadis,goldberg,chacko}. In this case, far from the
source, the massive radion field goes to zero exponentially, implying that the
volume of the supplementary space stays stable around the background value.

Therefore, to be consistent with observational data, extra-dimensional
theories, such as the ADD model, need an additional theoretical ingredient
that provides stabilization for the supplementary space volume. Beside this,
some models propose the existence of new self-interacting scalar fields that
inhabit the bulk \cite{BDextradim}. These scalars fields of Brans-Dicke type
will couple to the radion field, influencing its behavior. For these reasons,
it is interesting to treat $\phi$ and the tensor $h_{\mu\nu}$ as independent
fields from the phenomenological point of view and to investigate the supposed
effects of these fields separately on each experiment whenever is possible
\cite{graviscalars}.

Laboratory tests of the inverse square law of gravity by torsion-balance
experiments are capable of establishing constraints on each field individually
when some conditions are attained. For instance, if the Compton wavelength
$\lambda$ of the radion is greater than $R$, then, the experimental data put
bounds on the strength of the radion \cite{adelbergREV,radion}. On the other
hand, under the condition $\lambda<<R$, the tensor field is the quantity that
is constrained \cite{adelbergREV,radion}.

Another important test of the large extra dimensions theories comes from
high-energy colliders. It happens that, in these kind of experiments, the
fields $\phi$ and $h_{\mu\nu}$ are not probed with the same accuracy at the
tree level \cite{colliders}. The reason is that, according to (\ref{radion}),
the source term of the radion is the trace of the energy-momentum tensor.
Therefore, radiation or any pure relativistic source is not capable of
producing the field $\phi$. In fact, the strength of the radion is limited by
the rest mass of the particles when the source is on the mass-shell. Another
restrictive aspect of the radion-matter interaction is the fact that the
effects of the radion field on the motion of the particles are also limited by
their rest mass and tends to vanish as the particle's velocity approaches the
speed of light. Thus, in a collision with energy $E\text{,}$ the influence of
the field $\phi$, in comparison to the contributions of the tensor $h_{\mu\nu
}$, is reduced by a factor of the order of $\left(  m/E\right)  ^{2}$, where
$m$ is the rest mass of the heaviest particle involved in the collision
\cite{colliders}.

So, it is relevant to find alternative systems from which we can get new and
independent bounds for the radion field. In the atomic system, as the matter
is found in a non-relativistic regime, the fields can be tested at same level
of precision. Motivated by this idea, we intend to investigate the effects of
the radion field on the atomic energy spectrum of the muonic Helium. In this
system, the nucleus is the source of a gravitational field that is probed by
the muon, which plays the role of a test particle.

The behavior of a particle in curved spacetimes is dictated by the Lagrangian
$L=-mc\sqrt{-g_{AB}\dot{x}^{A}\dot{x}^{B}}\text{,}$ where $\dot{x}^{A}$ is the
particle's proper velocity. From this Lagrangian, we can find that, in the
weak field regime, the interaction of the particle with an external
gravitational field is given by $L_{I}=\frac{1}{2}m\hat{h}_{AB}\dot{x}^{A}%
\dot{x}^{B}$. In this order of approximation, this Lagrangian can be rewritten
as $L_{I}=\frac{1}{2}\hat{h}_{AB}P^{A}\dot{x}^{B}$, where $P^{A}=\partial
L/\partial\dot{x}^{A}$ is the conjugated momentum of the particle. Clearly,
the term $P^{A}\dot{x}^{B}$ can be interpreted as the flux of the particle's
momentum in spacetime. Thus, when we are dealing with fields, this term is
equivalent to the energy-momentum tensor $T^{AB}$ of the field, and therefore
the corresponding Lagrangian of interaction will be translated as $L_{I}%
=\frac{1}{2}\int\left(  \hat{h}_{AB}T^{AB}\right)  d^{3+\delta}X$, which
coincides with the expression obtained in \cite{colliders}.

Here we intend to focus our attention on the interaction mediated by the
radion. Considering the form of $T^{AB}$ for confined fields and the
decomposition of $\hat{h}_{\mu\nu\text{,}}$ we find that the coupling between
the radion and the matter is given by the Lagrangian:
\begin{equation}
L_{I}=\dfrac{1}{2}\int\phi T_{[t]}\left(  x\right)  f_{[t]}\left(  z\right)
d^{3+\delta}X \label{phiT}%
\end{equation}
where $T_{\left[  t\right]  }=\eta_{\mu\nu}T_{[t]}^{\mu\nu}$ is the trace of
the energy-momentum tensor of the test particle. We are using the $t$-index in
reference to the test particle's quantities. Now, using equation
(\ref{radion}), we can express the radion field in terms of $T_{\left[
N\right]  },$ i.e., the trace of the energy-momentum tensor of the nucleus.
Thus, it follows from (\ref{phiT}), that the behavior of the test particle
(i.e. the muon) under the gravitational influence of the nucleus mediated by
the radion field is described by the Lagrangian:
\begin{equation}
L_{I}=\dfrac{1}{2}\dfrac{\hat{G}_{D}}{c^{4}}\dfrac{\delta}{2\left(
\delta+2\right)  }\sum_{i}\int\int\dfrac{T_{\left[  N\right]  }\left(
x_{i}^{\prime}\right)  T_{\left[  t\right]  }\left(  x\right)  f_{\left[
N\right]  }\left(  z_{i}^{\prime}\right)  f_{\left[  t\right]  }\left(
z\right)  }{\left\vert \vec{X}-\vec{X}_{i}^{\prime}\right\vert ^{1+\delta}%
}d^{3+\delta}X_{i}^{\prime}d^{3+\delta}X\text{.} \label{L}%
\end{equation}
\qquad{}\qquad{}\qquad{}\qquad{}\qquad{}

In the non-relativistic regime, as the time-time component of the
energy-momentum tensor is much greater than the others, then $T_{\left[
t\right]  }\left(  x\right)  $ can be approximated by $c^{2}m_{\mu}%
\rho_{\left[  \mu\right]  }\left(  x\right)  $, where $\rho_{\left[
\mu\right]  }\left(  x\right)  $ is the normalized mass density of the muon,
which has a mass $m_{\mu}$. If we write the normalized mass density of the
muon in terms of its field, $\psi_{\left[  \mu\right]  }$, as $\rho_{\left[
\mathbf{\mu}\right]  }=\psi_{\left[  \mathbf{\mu}\right]  }^{\dag}%
\psi_{\left[  \mathbf{\mu}\right]  }$, then we can find, from (\ref{phiT}),
that the associated Hamiltonian is simply $H_{I}=-L_{I}$ in this regime.

Therefore, the Hamiltonian of the gravitational interaction between the
nucleus and the muon through a massless radion field can be written, in this
order of approximation, as:
\begin{equation}
H_{I}=-\frac{\hat{G}_{D}m_{\mu}\delta}{4c^{2}\left(  \delta+2\right)  }%
\sum_{i}\int\int\frac{T_{\left[  N\right]  }\left(  x^{\prime}\right)
\rho_{\lbrack\mu]}\left(  x\right)  f_{\left[  N\right]  }\left(
z_{i}^{\prime}\right)  f_{\left[  \mu\right]  }\left(  z\right)  }{\left\vert
\vec{X}-\vec{X}_{i}^{\prime}\right\vert ^{1+\delta}}d^{3+\delta}X_{i}^{\prime
}d^{3+\delta}X\text{.} \label{H}%
\end{equation}
In the case of a massive radion, a decreasing exponential factor, such as
$\eta e^{-r/\lambda}$, should be considered in the above integral. In this
exponential, the constant $\lambda$ is the radion Compton wavelength and the
adimensional constant $\eta$ measures any modification of the radion-matter
gravitational coupling that the stabilization mechanism could introduce.

The influence of the radion field in the atomic energy levels can be computed
from the average value $\left\langle H_{I}\right\rangle $ of the Hamiltonian
(\ref{H}) in the atom's states. If we consider that the radion's Compton
wavelength is greater than the nuclear radius, then the most stringent
constraints for the radion interaction at short distances can be extracted
from transitions involving the $S$-level, due to the overlapping between the
wave-functions of the muon and nucleus. It happens that the influence of the
gravitational interaction between the muon and the Helium nucleus on these
levels cannot be calculated in the infinitely thin brane scenario when
$\delta>1$, since, as pointed out in Ref. \cite{lemos1}, the internal
gravitational potential inside the nucleus is not computable when the
functions $f\left(  z\right)  $ are idealized as delta-like distributions.

This difficulty can be circumvented in thick brane scenarios, where the brane
has a characteristic width and the transverse profiles $f$ are regular
distributions concentrated inside the brane. In the leading order, we find
\begin{equation}
\left\langle H_{I}\right\rangle _{S}=-\frac{\eta\hat{G}_{D}m_{N}m_{\mu}\delta
}{4\left(  \delta+2\right)  \varepsilon^{\delta-2}}\left\vert \psi
_{S}(0)\right\vert ^{2}, \label{Hint}%
\end{equation}
where $\psi_{S}(0)$ is the wavefunction of the muon evaluated in the center of
the nucleus, $m_{N}$ is the nuclear mass that follows from the integration of
$T_{\left[  N\right]  }$ and the parameter $\varepsilon$ is a kind of an
effective distance in the transversal directions between the nucleus and muon
defined by the expression:
\begin{equation}
\frac{1}{\varepsilon^{\delta-2}}=\Gamma\left(  \delta/2\right)  \int
\frac{f_{N}\left(  z^{\prime}\right)  f_{t}\left(  z\right)  }{\left\vert
z-z^{\prime}\right\vert ^{\delta-2}}d^{\delta}z^{\prime}d^{\delta}z.
\label{epsilon}%
\end{equation}
The value of this effective distance depends on the overlapping of the
transversal profiles of the confined fields. When both transversal functions
are identical normal distributions, the parameter $\varepsilon$ is equal to
the distribution's standard deviation multiplied by two.

The gravitational constant $G_{D}$, defined in the ambient space, establishes
a new length scale $\ell_{D}^{\delta+2}=G_{D}\hbar/c^{3}$. Therefore, the
energy shift on $S$-level due to the gravitational interaction depends,
according to expression (\ref{Hint}), on an effective length scale defined by
$\ell_{eff}^{4}=\ell_{D}^{\delta+2}/\varepsilon^{\delta-2}$ or, equivalently,
on the effective energy scale $\Lambda=hc/\ell_{eff}.$ Writing wave functions
of the $nS$-level in terms of the Bohr radius of the muonic Helium ion, we
find that
\begin{equation}
\left\langle H_{I}\right\rangle _{nS}=-c^{7}h^{3}\frac{m_{N}m_{\mu}}%
{\Lambda_{r}^{4}}\frac{4\pi}{n^{3}[a_{0}(\mu^{4}He^{+})]^{3}}, \label{Hs}%
\end{equation}
where the effective energy scale of the radion $\Lambda_{r}$ is defined in
terms of $\Lambda$ by absorbing the $\delta$-dependent factor and the
enhancing factor $\eta,$ according to the expression:
\begin{equation}
\frac{1}{\Lambda_{r}^{4}}=\frac{\delta}{\left(  \delta^{2}-4\right)
\pi^{\delta/2}}\frac{\eta}{\Lambda^{4}}.
\end{equation}
This expression is valid for $\delta>2$.

The gravitational shift of the energy of the $2P$-level is weaker by a factor
of the order of ($a_{0}/R_{\alpha})^{2}$ in comparison to (\ref{Hs}) and can
be neglected in the first approximation. Thus, the gravitational interaction
between the muon and the alpha particle will increase the difference between
the $2P_{1/2}$ and $2S$ level by the amount $\Delta E_{G}=-\left\langle
H_{I}\right\rangle _{2S}$.

\section{Constraints for the alpha particle radius and radion's effective
energy scale}

Recent measurements of the $2S-2P$ transition in the muonic helium-$4$ ion
have tried to shed light on the proton radius puzzle. From these precise
transition data, it is possible to extract the root-mean-square charge radius,
$r_{\alpha}$, of the $\alpha$ particle with high precision \cite{nature2}. The
new value is compatible with the charge radius obtained from scattering
experiments between the electron and $^{4}He$ \cite{scatt}, unlike what
happens with analogous measurements involving muonic hydrogen and deuterium.
Therefore, we can use these measurements to put new bounds on parameters of
non-standard physics theories.

The energy difference between the $2P_{1/2}$ and $2S$ levels of the $\left(
\mu^{4}He\right)  ^{+}$ can be calculated with great accuracy based on the
Standard Model. According to Ref. \cite{nature2}, it is given by:
\begin{equation}
\Delta E_{\left(  2P_{1/2}-2S\right)  }=\left[  1677.690-106.220\times\left(
\frac{r_{\alpha}^{2}}{\text{fm}^{2}}\right)  \right]  \text{meV},
\end{equation}
where the first term has an uncertainty of 0.292 meV and, in the second term,
the uncertainty of the numeric coefficient is 0.008 meV.

In the thick brane scenario, the gravitational interaction will increase the
gap between these levels. According to the calculation of the last section, in
the leading order, the previous expression would contain a new term that
depends on the unknown parameter $\Lambda_{r}$:
\begin{equation}
\Delta E_{\left(  2P_{1/2}-2S\right)  }=\left[  1677.690-106.220\times\left(
\frac{r_{\alpha}^{2}}{\text{fm}^{2}}\right)  +5.182\times10^{-7}\times\left(
\frac{\text{TeV}}{\Lambda_{r}}\right)  ^{4}\right]  \text{meV}.\label{E_theo}%
\end{equation}
In this equation, $\Lambda_{r}$ is expressed in TeV units and its numeric
factor is calculated from equation (\ref{Hs}), by using the CODATA recommended
values for that quantities. The theoretical prediction (\ref{E_theo}) should
be compared with the experimental value $\Delta E_{\left(  2P_{1/2}-2S\right)
}^{\exp}=\left(  1378.521\pm0.048\right)  $meV \cite{nature2}.

Let us admit that the theoretical and experimental values should coincide
within the combined uncertainty $\delta E=\left(  \delta E_{th}^{2}+\delta
E_{\exp}^{2}\right)  ^{1/2}$, i.e., $\Delta E_{\left(  2P_{1/2}-2S\right)
}=\Delta E_{\left(  2P_{1/2}-2S\right)  }^{\exp}\pm\delta E$. This condition
determines constraints that must be satisfied by $\Lambda_{r}$ and $r_{\alpha
}$ together. The shadow regions in Figure 1 correspond to the values of the
parameters $\left(  r_{\alpha},\Lambda_{r}\right)  $ permitted by the data.
The inner and darker area corresponds to regions at 68\% confidence level. The
wider region has a 95\% confidence level. \begin{figure}[ptb]
\includegraphics[scale=0.4]{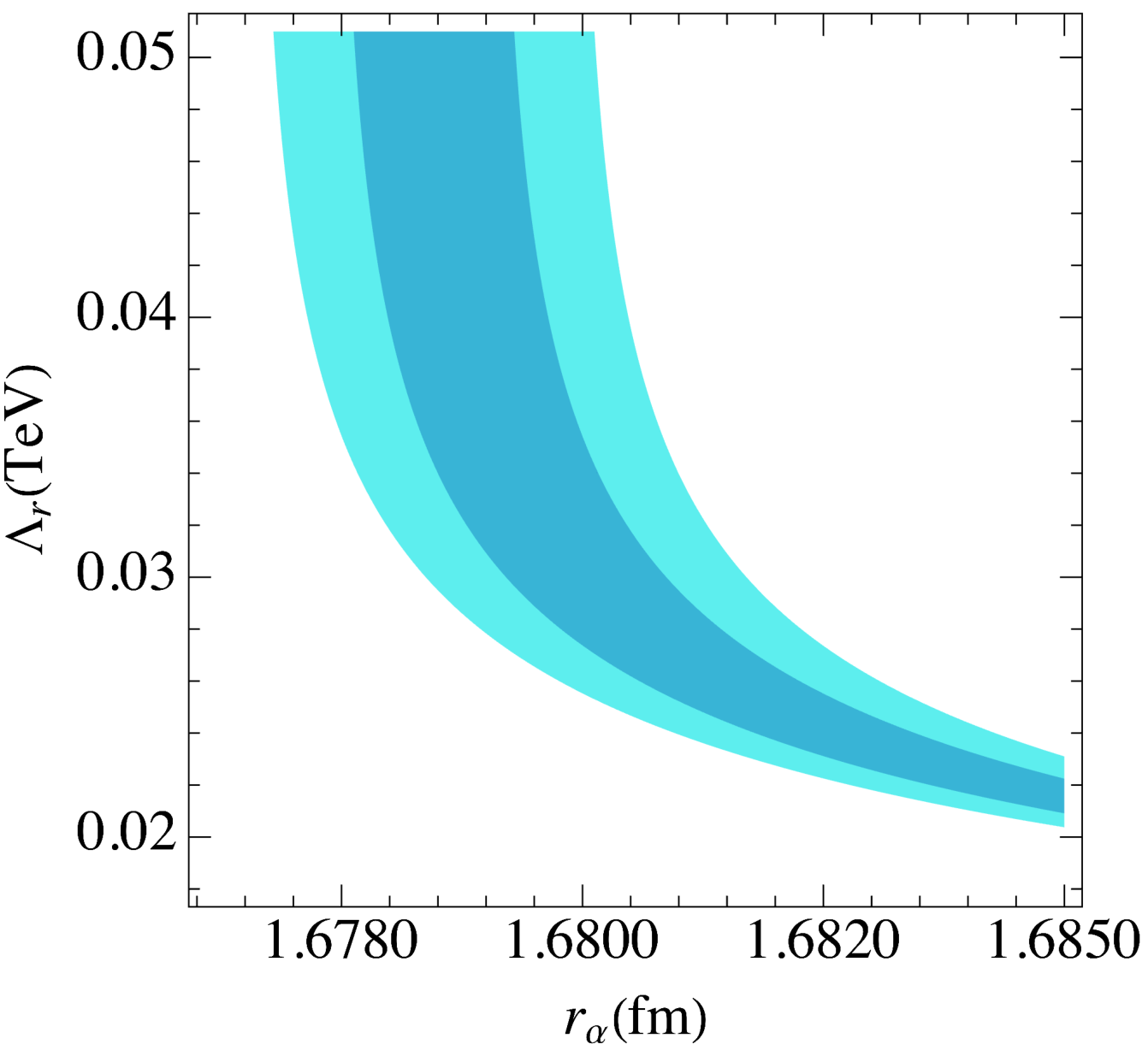}
\par
\label{fig1}\caption{Constraints for the effective energy scale ($\Lambda_{r}%
$) of the radion and the alpha-particle charge radius ($r_{\alpha}$)
established from recent data of 2P-2S transitions in the muonic Helium-4 ion.}%
\end{figure}

The allowed regions are compatible with the absence of extra dimensions
($\Lambda_{r}\rightarrow\infty$). In this case, the charge radius of the
alpha-particle would be $r_{\alpha}=1.67824(83)$fm. Figure\nolinebreak\ 1
explores a domain where alpha-particle radius is in a range delimited from the
\textit{e-He} scattering, namely, $r_{\alpha}(scatt)=1.681\pm0.004$ fm
\cite{scatt}. According to the data, the strongest effects of
extra-dimensional occur for greater alpha radius $r_{\alpha}$.

To obtain an estimate for the maximum contribution of the radion to the energy
gap between $2P_{1/2}$ and $2S$ states, let us consider $r_{\alpha}=1.689$ fm,
i.e., the value in the limit of 2 error-bars of the \textit{e-He} scattering
experiment. In this case, the energy from extra-dimensional gravitational
interaction between the alpha nucleus and muon can reach 4.4 meV,
corresponding to an effective energy scale of about 18 GeV, found at the
border of the 2$\sigma$-confidence level region.

\section{The proton radius puzzle}

As we have already mentioned, the size of the alpha particle measured by using
muonic spectroscopy agrees with the radius inferred from the scattering of
electrons by helium. Moreover, the proton radius obtained from recent
measurements of hydrogen atom transitions is compatible with the value derived
from muonic spectroscopy \cite{hessels}. These results have reinforced the
idea that, probably, it is not necessary to resort to nonstandard physics to
explain the proton radius puzzle \cite{hessels,canada,solution}.

However, as a definitive explanation is not completely known yet, it is
interesting to investigate all possibilities \cite{jentchura}. In this section
we intend to check the implications of the constraints found here on this
issue. More specifically, we want to examine whether the higher-dimensional
gravity could be excluded as a possible explanation for the puzzle based on
the present data.

In the thick brane scenario, the Lamb shift of muonic hydrogen has a new
contribution coming from the proton-muon gravitational interaction in higher
dimensions. To solve the radius puzzle, the additional energy separation
between $2P_{1/2}$ and $2S$ levels produced by this interaction should be of
the order of $0.3$ meV, the unexpected excess of energy found experimentally
in the Lamb shift of muonic hydrogen. It happens that, according to the data
considered here, $\Lambda_{r}=18$ Gev is the effective energy scale of the
radion which gives the strongest effect of extra dimensions on the energy
levels of the atom and, at the same time, is compatible with the scattering
radius of the alpha particle obtained from the e-He interaction. However, with
this energy scale, the gravitational interaction between the proton and muon,
mediated by the radion, would increase the Lamb shift of the muonic hydrogen
just by $0.1$ meV approximately.

At first sight, we are led to think that higher-dimensional gravity could be
discarded as an explanation for the puzzle based on this result. However, we
should keep in mind that the effective scale $\Lambda_{r}$ depends on $G_{D}$,
$\eta,$ and, in particular, on the parameter $\varepsilon$, defined by
equation (\ref{epsilon}), which can be interpreted as an effective distance
between the muon and nucleus in the supplementary space. As this effective
transverse distance depends on the overlapping of the functions $f_{muon}%
\left(  z\right)  $ and $f_{N}\left(  z\right)  ,$ which describes,
respectively, the energy-momentum distribution of the muon and nucleus in the
extra space, then the parameter $\varepsilon$ could, in principle, assume
different values for distinct muonic atoms.

The shorter the parameter $\varepsilon$, the stronger the effect. If the
radion exchange in the muonic helium can enlarge the energy difference between
$2P_{1/2}$ and $2S$ states by $4.4$ meV, then the radion would be capable of
increasing the Lamb shift of $\mu H$ by $0.3$ meV, as long as the effective
transversal distance between the muon and proton in $\mu H$ be shorter than
the transversal distance between the muon and the alpha nucleus in $\mu
^{4}He^{+}$. Indeed, considering equation (\ref{Hs}) and expressing
$\Lambda_{r}$ in terms of $\varepsilon$, we can find that the gravitational
potential energy of the muonic hydrogen and of muonic helium both in 2S-state
obey the following relation:\qquad{}
\begin{equation}
\frac{\left.  \left\langle H_{I}\right\rangle _{2S}\right\vert _{\mu H}%
}{\left.  \left\langle H_{I}\right\rangle _{2S}\right\vert _{\left(  \mu
^{4}He\right)  ^{+}}}=\frac{m_{H}}{m_{\alpha}}\left(  \frac{a_{0}(\mu
^{4}He^{+})}{a_{0}(\mu H)}\right)  ^{3}\left(  \frac{\varepsilon_{\mu
^{4}He^{+}}}{\varepsilon_{\mu H}}\right)  ^{\delta-2} \label{Relative_energy}%
\end{equation}

If the parameter $\varepsilon$ had the same value in the two atoms, the effect
of the radion on the energy of the $S$-states of muonic hydrogen would be
almost 40 times smaller when compared with its influence on the same states of
muonic helium.

According to equation (\ref{Relative_energy}), to get $\left.  \left\langle
H_{I}\right\rangle _{2S}\right\vert _{\mu H}$ $=-0.3$ meV and $\left.
\left\langle H_{I}\right\rangle _{2S}\right\vert _{\left(  \mu^{4}He\right)
^{+}}$ $=-4.4$ meV, the muon-proton transversal distance in hydrogen atom
should be shorter than the transversal distance between the muon and helium
nucleus by the following factor:
\begin{equation}
\frac{\varepsilon_{\mu H}}{\varepsilon_{\mu^{4}He^{+}}}\simeq\left(  \frac
{1}{2.74}\right)  ^{1/(\delta-2)}. \label{relative_distance}%
\end{equation}

Some versions of braneworld models assume that quarks could be localized in
different slices of a thick brane \cite{thickbrane}. Therefore, distinct
quarks could be at different transversal distances from the muon when they are
in a bound state, forming an atom. Equation (\ref{Relative_energy}) is, in
this sense, connected to these types of models since it takes this possibility
into account. However, the parameter $\varepsilon$ depends on the nucleon's
mass distribution, which relies mostly on the energy of quark-gluon
interactions. Therefore, a theoretical justification of relation
(\ref{relative_distance}) demands further investigations on the question of
the energy distribution of quark-gluon interactions in the transversal directions.

\section{Concluding Remarks}

A very interesting phenomenological implication of braneworld theories is the
strengthening of the gravitational interaction over short distances. In this
context, muonic atoms arise as promising systems for testing supposed
modifications of gravity at the atomic scale. Because the mass of the muon is
207 times greater than the electron's mass, muonic atoms are more sensitive to
probe deviation in the gravitational potential than conventional atoms.
Indeed, when a muonic hydrogen is in an $S$ -state, the energy of the
proton-muon gravitational interaction in a thick brane model is more than 2
billion times greater than the energy of the same interaction acting within
the conventional hydrogen. In this paper, we exploited this sensitivity to
test the predictions of braneworld theories by using new experimental data on
the Lamb shift of the muonic helium-4 ion.

More specifically we have calculated explicitly the effects of the radion, the
scalar degree of freedom of the higher-dimensional gravity, on the energy
difference between the $2P_{1/2}$ and $2S$ levels of the $\left(  \mu
^{4}He\right)  ^{+}$ . According to the thick braneworld, the gravitational
interaction between muon and helium-4 nucleus will increase the gap between
these levels by an amount that depends on an effective energy scale of the
radion $\Lambda_{r}$ . This quantity is a free parameter of the model defined
in terms of $G_{D}$ (the gravitational constant of the higher-dimensional
space), $\eta$ (the enhancing factor the radion field could acquire from a
stabilization mechanism, for instance) and $\varepsilon$ (an effective
transversal distance between the muon and nucleus of the helium). By using the
spectroscopic data of muonic helium-4 ions, which are consistent with the
radius of the alpha particle inferred from the electron-helium scattering
experiment, we have find experimental bounds for $\Lambda_{r}$ as function of
$r_{\alpha}$.

According to the available data, the strongest extradimensional effect the
radion is capable to produce on the Lamb shift of $\left(  \mu^{4}He\right)
^{+}$ is a 4.4 meV increase. From this result, we can deduce what is the
additional separation energy in the Lamb shift of the $\mu H$ caused by the
gravitational interaction between the proton and muon mediated by the radion.
From the equation (\ref{Relative_energy}), we see that, comparatively, the
effect on the energy of S-levels produced by the radion exchange in the two
muonic atoms depends on the ratio between the transverse distances
$\varepsilon_{\mu^{4}He^{+}}$ and $\varepsilon_{\mu H}$. When the effective
transversal distance between the muon and proton in $\mu H$ atom is
sufficiently short (see equation (\ref{relative_distance})), the radion can
account for the unexplained excess of energy of 0.3 meV in the Lamb shift of
the muonic hydrogen. In the case of six extra dimensions, for instance,
$\varepsilon_{\mu H}$ should be about 78\% of $\varepsilon_{\mu^{4}He^{+}}$.
The effective transverse distance depends on the profile in the supplementary
space of the energy-momentum distribution of the muon and atomic nucleus. Some
braneworld models predict that distinct quarks are stuck in different slices
of a thick brane \cite{thickbrane}, therefore their distances from the muon
would be different. However, a theoretical justification of the relation
(\ref{relative_distance}) demands further investigations on the energy
distribution of the quark-gluon interaction (that provides the most part of
nucleus mass) in the transverse direction of the 3-brane in this scenario.

\begin{acknowledgments}
ASL acknowledge support from CAPES (Grant no. 88887.800922/2023-00).
\end{acknowledgments}

\end{document}